\documentclass[aps,prb,10pt,twocolumn,floatfix,superscriptaddress,showpacs,numerical,footinbib,groupedaddress]{revtex4-1}

\usepackage{graphicx}
\usepackage{amsmath}
\usepackage{amssymb}
\usepackage{hyperref}
\usepackage{soul}
\usepackage{bbold}

\newcommand{\im}{\mathrm{i}}

\newcommand{\intracellular}{\textrm{intra-cellular$~$}}
\newcommand{\intercellular}{\textrm{inter-cellular$~$}}

\begin{document}

\title{Bulk-boundary correspondence from the inter-cellular Zak phase}

\begin{abstract}

The Zak phase $\gamma$, the generalization of the Berry phase to Bloch wave functions in solids, is often used to characterize inversion-symmetric 1D topological insulators;
however, since its value can depend on the choice of real-space origin and unit cell, only the difference between the Zak phase of two regions is believed to be relevant.
Here, we show that one can extract an origin-independent part of $\gamma$, the so-called inter-cellular Zak phase $\gamma^{\mathrm{inter}}$, which can be used as a bulk quantity to predict the number of surface modes as follows:
a neutral finite 1D tight-binding system has $n_s = \gamma^{\mathrm{inter}}/\pi$ (mod 2) number of in-gap surface modes below the Fermi level if there exists a commensurate bulk unit cell that respects inversion symmetry.
We demonstrate this by first verifying that $\pm e\gamma^{\mathrm{inter}}/2\pi$ (mod $e$) is equal to the extra charge accumulation in the surface region for a general translationally invariant 1D insulator, while the remnant part of $\gamma$, the intra-cellular Zak phase $\gamma^{\mathrm{intra}}$, corresponds to the electronic part of the dipole moment of the bulk's unit cell.
Second, we show that the extra charge accumulation can be related to the number of surface modes when the unit cell is inversion symmetric.
This bulk-boundary correspondence using $\gamma^{\mathrm{inter}}$ reduces to the conventional one using $\gamma$ when the real-space origin is at the inversion center.
Our work thereby clarifies the usage of $\gamma$ in the bulk-boundary correspondence.
We study several tight binding models to quantitatively check the relation between the extra charge accumulation and the inter-cellular Zak phase as well as the bulk-boundary correspondence using the inter-cellular Zak phase.
\end{abstract}

\author{Jun-Won \surname{Rhim}}
\affiliation{Max-Planck-Institut f{\"u}r Physik komplexer Systeme, 01187 Dresden, Germany}

\author{Jan \surname{Behrends}}
\affiliation{Max-Planck-Institut f{\"u}r Physik komplexer Systeme, 01187 Dresden, Germany}

\author{Jens H. \surname{Bardarson}}
\affiliation{Max-Planck-Institut f{\"u}r Physik komplexer Systeme, 01187 Dresden, Germany}

\maketitle

\section{Introduction}
\label{sec:introduction}

The Berry phase is a geometric phase of eigenstates obtained when cyclically varying external parameters.\cite{Berry1984}
For a Hamiltonian $\mathcal{H}(\mathbf{R})$ that depends on the external parameters $\mathbf{R}$, the Berry phase is given by
\begin{equation}
	\gamma_n = \im\oint_C \langle n(\mathbf{R})|\nabla_\mathbf{R}| n(\mathbf{R})\rangle \cdot d\mathbf{R},
\end{equation}
where $| n(\mathbf{R})\rangle $ is the $n$-th eigenvector and $C$ is a closed loop in the parameter space of $\mathbf{R}$.
In a Brillouin zone, as pointed out by Zak\cite{Zak1989}, a natural choice for the cyclic parameter is the crystal momentum $k$.
The explicit form of the resulting Zak phase of the $n$-th band is 
\begin{equation}
	\gamma_n = \im\int_\mathrm{BZ}dk
	\langle u_{n,k}|\partial_k|u_{n,k}\rangle,
	\label{eq:Zak_n}
\end{equation} where BZ represents the 1D Brillouin zone, and $u_{n,k}=\sqrt{N}e^{-\im kx}\psi_{n,k}$ is the periodic part of the Bloch function $\psi_{n,k}$.
The total Zak phase $\gamma$ is obtained by summing $\gamma_n$ over filled bands.

The Zak phase was endowed with a physical meaning with the modern definition of the polarization $\mathbf{P}$, introduced by Vanderbilt and King-Smith.\cite{Vanderbilt1993a,Vanderbilt1993b}
The ambiguity of the classical polarization, which depends on the shape of the unit cell boundary,\cite{Resta1994} was resolved by redefining the electronic part of the polarization $\mathbf{P}^{\text{el}}$ as an integral of Wannier functions over the whole space, rather than over a unit cell.
With this definition, one can accurately predict the bound surface charge, $\sigma = \mathbf{P}\cdot\hat{n}$ with $\hat{n}$ the surface orientation, measured in capacitance experiments.\cite{Spaldin2012}
Furthermore, the electronic polarization $P^{\text{el}}_\perp$ along $\hat{n}$ is related to the Zak phase (evaluated along $\hat{n}$ with $\mathbf{k}_\parallel$ kept fixed) via
\begin{equation}
	P^{\text{el}}_\perp = \mathbf{P}^{\text{el}}\cdot\hat{n} = -\frac{e}{(2\pi)^3}\int_\mathcal{A}d\mathbf{k}_\parallel\sum_{n=1}^Z\gamma_n,
\end{equation}  
where $\mathcal{A}$ is the 2D Brillouin zone projected to the 2D plane perpendicular to $\hat{n}$, $\mathbf{k}_\parallel$ is the momentum component in this plane, and $Z$ the number of filled bands.

In the currently active study of topological aspects of materials,\cite{Kane2005a,Kane2005b,Bernevig2006,Fu2007a,Fu2007b,Hasan2010,Xiao-Liang2011,Liang2011,Shunji2013} the Zak phase has been utilized as a topological number to classify various genuine 1D topological insulators, as well as effective ones, such as those obtained by fixing one or two momenta of 2D or 3D Hamiltonians;\cite{Hatsugai2006,Ryu2006,Delplace2011, Hatsugai2013,Grusdt2013,Barnett2013,Chan2015,Vanderbilt2011,Hodge2011,Atala2013,Longhi2013,Budich2013,Yoshimura2014,Bernevig2014,Fernando2014,Ling2015,Anna2015,Lee2015,Xiao2016}
it was naturally extended to the concept of non-abelian Wilson loops in the multi-band case and used for classifications of topological insulators with inversion or non-symmorphic symmetries and topological crystalline insulators.\cite{Fidkowski2011,Bernevig2014b,Vanderbilt2014,Bernevig2016a,Bernevig2016b,Bernevig2016c}
Furthermore, the Zak phase has been widely used for the Z$_2$ classification of inversion symmetric 1D systems where it is quantized to $0$ or $\pi$ (mod $2\pi$).\cite{Zak1989}
In this case, the conventional bulk-boundary correspondence states that there are boundary modes if the Zak phase is nontrivial, $\gamma = \pi$, while $\gamma = 0$ is considered a trivial insulator without surface modes.\cite{Ryu2006, Delplace2011, Hatsugai2013, Grusdt2013, Barnett2013, Chan2015}

Recent work, however, pointed out that the Zak phase depends on the gauge choice of choosing the origin of the real space, and how one defines boundaries of the unit cell, although it is invariant under gauge transformation of the form $u_{n,k}\rightarrow e^{\im\phi_k}u_{n,k}$.\cite{Atala2013,Fernando2014,Ling2015}
This means that the Zak phase itself is not a well-defined topological number since it cannot characterize the bulk uniquely.
In an attempt to resolve this ambiguity, Atala et al.\cite{Atala2013} suggested that the difference of the Zak phase between different states could be a proper topological number, and Juan et. al.\cite{Fernando2014} revised the Zak phase by adding a unit cell dependent term such that the resultant Z$_2$ number plays the role of a gauge invariant topological number.
In spite of these issues, the conventional bulk-boundary correspondence using the Zak phase has been successfully applied in many cases\cite{Ryu2006, Delplace2011, Hatsugai2013, Grusdt2013, Barnett2013, Chan2015}.
Furthermore, additional conditions for the applicability of the correspondence to finite systems have been given, such as that terminated edges should not break the inversion symmetry of the bulk \cite{Fu2015,Murakami2016} or that the finite system should be commensurate with the bulk's unit cell\cite{Fernando2014}.
However, the necessity of those assumptions has never been demonstrated in general 1D systems, and we even find a counterexample for the former one in Sec.~\ref{sec:examples}~B where the conventional bulk-boundary correspondence using $\gamma$ does not work even in the presence of inversion symmetry both in the bulk and the terminated system.

\begin{figure*}
\includegraphics[width=1.7\columnwidth]{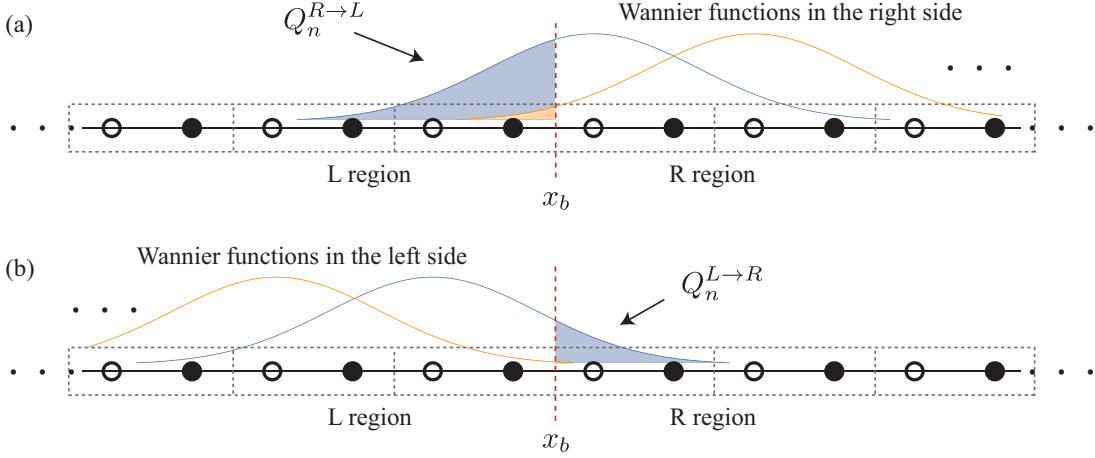}
\caption{(Color online) A schematic figure describing a general 1D system and its Wannier functions. Gray dashed boxes are the unit cells, which contain atomic sites marked by filled and empty circles as an example. Each unit cell can in general have any number of atomic sites. The 1D bulk is divided into left and right sides with respect to a given unit cell boundary $x_b$, which is marked by the red dashed line. We call these regions the $L$ and $R$ region. Wannier functions belong to the left or right side according to their centers. $Q^{R\rightarrow L}_n$ represents how much weight of right side's Wannier functions are in the left side, and vice versa for $Q^{L\rightarrow R}_n$. 
%Here, $n$ is the band index and we sum over occupied bands to obtain total weights of them.
}
\label{fig:left-right}
\end{figure*}

In this paper, we clarify these issues by providing a more detailed analysis of the Zak phase, concentrating on general 1D tight-binding models.
To this end, we split the Zak phase into two terms, the \textit{intra-cellular} and \textit{inter-cellular} Zak phase (this splitting was earlier introduced by Kudin et al.\cite{Resta2007a}), and provide them with their proper physical interpretations.
The intra-cellular Zak phase $\gamma^{\mathrm{intra}}$ describes the electronic part of the classical polarization of the bulk's unit cell, and the inter-cellular Zak phase $\gamma^{\mathrm{inter}}$ represents the difference between the net weight of the Wannier functions in the left and right sides of the 1D system with respect to a unit cell boundary, with their centers belonging to opposite sides as illustrated in Fig.~\ref{fig:left-right}.
We further show that, if the system is terminated, this interpretation of the inter-cellular Zak phase leads to an accurate prediction of the extra charge accumulations in a region around the surface.
We assume that these surface regions are commensurate with the bulk's unit cell, and large enough that effects of the termination are negligible deep inside the bulk.
Here, the extra charge accumulation is just the total charge in the surface region including ionic contributions. 
Note that it is a different quantity from the \textit{bound surface charge} introduced by Vanderbilt et al.\cite{Vanderbilt1993b}, which can be measured in capacitance experiments.
For instance, the bound surface charge can be nonzero even when the surface regions remain neutral while the extra charge accumulation vanishes in this case.
Although we have obtained a general relation between $\gamma^\mathrm{inter}$ and the extra charge accumulation around surfaces for any translationally invariant systems, it cannot generally be used to count the number of surface modes in the gap below the Fermi level, since it includes charge densities both from bulk bands and surface states; however, we show that it is possible if inversion symmetry is respected in the unit cell.
In this case, there are $n_s = \gamma^{\mathrm{inter}}/\pi$ (mod 2) surface modes below the Fermi level in a finite system, if it is commensurate with an inversion symmetric unit cell of the bulk and $\gamma^{\mathrm{inter}}$ is evaluated from this commensurate unit cell.   
We can replace $\gamma^{\mathrm{inter}}$ with $\gamma$ if the real-space origin is chosen to be located at the inversion center of the unit cell because $\gamma^{\mathrm{intra}}$, an origin-dependent quantity, vanishes in this case.
In other words, the conventional bulk-boundary correspondence using the total Zak phase is valid if there exist an inversion symmetric bulk unit cell commensurate with the finite system and the real-space origin is at the inversion center.

The rest of the paper is organized as follows:
in Sec.~\ref{sec:reinterpretation} and \ref{sec:extra_charge} we provide the physical meaning of the \textrm{intra-} and \textrm{inter-cellular} Zak phase and show how they are related, respectively, to the electronic part of the classical polarization of the bulk and the extra charge accumulation in the surface region.
In Sec.~\ref{sec:bulk_boundary}, based on this physical interpretation of the parts of the Zak phase, we develop a reformulated bulk-boundary correspondence of one-dimensional systems employing the \textrm{inter-cellular} Zak phase.
In Sec.~\ref{sec:examples} we apply our interpretation of the inter-cellular Zak phase and the bulk-boundary correspondence to the Rice-Mele 1D-chain model and two 2D toy models which can be treated as effective 1D systems by fixing one of the momenta.
Finally, discussions and concluding remarks are given in Sec.~\ref{sec:conclusions}.

%%%%%%%%%%%%%%%%%%%%%%%%%%%%%%%%%%%%%%%%%%%%%%%%%%%%%%%%%%
%%%%%%%%%%%%%%%%%%%%%%%%%%%%%%%%%%%%%%%%%%%%%%%%%%%%%%%%%%

\section{Intra- and inter-cellular Zak phase}
\label{sec:reinterpretation}

Let us consider a general translationally-invariant one-dimensional system described within the tight-binding approximation.
The system consists of $N_\mathrm{b}$ atomic sites per unit cell, each with $N_{\mathrm{orb}}^i$  orbitals, including the spin degree of freedom for spinful systems. 
The ionic charge at site $i$ is $Z_ie$, and the total ionic charge in the unit cell is $Ze=\sum_{i=1}^{N_\mathrm{b}}Z_ie$.
In order for the insulating system to be neutral, there are then necessarily $Z$ fully filled bands below the Fermi level.
Within the Born-von Karman boundary condition, $\psi(x+Na) = \psi(x)$, the Bloch eigenfunctions of this system take the form
\begin{eqnarray}
\psi_{n,k}(x) = \frac{1}{\sqrt{N}}\sum_{m}^N\sum_{i=1}^{N_\mathrm{b}}\sum_{\zeta=1}^{N_{\mathrm{orb}}^i} \alpha^{n,i,\zeta}_k \phi^{i,\zeta}_{m}(x) e^{\im kma}\label{eq:bloch}
\end{eqnarray}
where $N$ is the number of unit cells, and $m$ and $n$ are the unit cell and band index.
Here,
\begin{eqnarray}
\phi^{i,\zeta}_{m}(x) = \phi^{\zeta}(x-ma-b_ia)
\end{eqnarray}
is the $\zeta$-th atomic orbital centered at $ma+b_ia$ which is the position of the $i$-th atomic site in the $m$-th unit cell.
Atomic orbitals at the same site are orthonormal and the overlaps between orbitals on different sites are assumed to be exponentially vanishing in accordance with the tight-binding condition.\cite{note1}
The coefficients $\alpha^{n,i,\zeta}_k$ are obtained from solving the eigenvalue problem with the tight-binding Hamiltonian.
We choose the gauge in which every atomic orbital in the $m$-th unit cell has the same phase factor $e^{\im kma}$ in (\ref{eq:bloch}); the coefficients $\alpha$ then satisfy the periodic boundary condition $\alpha^{n,i,\zeta}_{k+G} = \alpha^{n,i,\zeta}_k$,  where $G$ is a reciprocal lattice vector.\cite{Resta2007a}
In another widely used gauge, the phase factor for the orbitals at $ma+b_ia$ is instead given by $e^{\im k(ma+b_ia)}$.\cite{Atala2013,Bernevig2014b,Chan2015,Bernevig2016b,Bernevig2016c}
In this case, the same Bloch eigenfunction $\psi_{n,k}(x)$ is obtained by replacing the coefficient $\alpha^{n,i,\zeta}_k$ with $\tilde{\alpha}^{n,i,\zeta}_k = \alpha^{n,i,\zeta}_ke^{-\im kb_ia}$, which satisfies the twisted boundary condition $\tilde{\alpha}^{n,i,\zeta}_{k+G}=\tilde{\alpha}^{n,i,\zeta}_ke^{-\im Gb_ia}$.
While both gauges yield the same Zak phase, since they describe the same Bloch eigenfunction $\psi_{n,k}(x)$, the former allows for a natural separation of the Zak phase into the intra- and inter-cellular part.

The Zak phase, following Kudin et al.\cite{Resta2007a}, can be split into two terms.
The $n$-th band Zak phase~\eqref{eq:Zak_n} is obtained from the lattice periodic part of the Bloch function
\begin{eqnarray}
u_{n,k}(x) &=& \sqrt{N}e^{-\im kx} \psi_{n,k}(x) \label{eq:periodic bloch}
\end{eqnarray}
which satisfies the orthonormality condition $\langle u_{n,k}| u_{n,k^\prime}\rangle = \int_\Omega dx u_{n,k}(x)^*u_{n,k^\prime}(x) = \delta_{k,k^\prime}$.\cite{Zak1989}
The inner product $\langle u_{n,k}|\partial_k u_{n,k}\rangle$ in the Zak phase is defined as
\begin{eqnarray}
\langle u_{n,k}|\partial_k u_{n,k}\rangle = \int_\Omega dx\; u^*_{n,k}(x)\frac{\partial}{\partial k}u_{n,k}(x),
\end{eqnarray}
with $\Omega$ a unit cell.
Then, the Zak phase is split into the intra-  and \intercellular Zak phase as
\begin{eqnarray}
\gamma_n = \gamma_n^{\mathrm{intra}} + \gamma_n^{\mathrm{inter}}
\end{eqnarray}
where
\begin{eqnarray}
\gamma_n^{\mathrm{intra}}  = N\int_{\mathrm{BZ}} dk \int_\Omega dx~ x \left| \psi_{n,k}(x) \right|^2 -2\pi m_\Omega, \label{eq:intra cellular zak}
\end{eqnarray}
where $m_\Omega$ is the index of the unit cell $\Omega$, and
\begin{eqnarray}
\gamma_n^{\mathrm{inter}}  = \im\sum_{i=1}^{N_b}\sum_{\zeta=1}^{N_{orb}^i} \int_{\mathrm{BZ}} dk \alpha^{n,i,\zeta *}_k \frac{\partial}{\partial k}\alpha^{n,i,\zeta}_k.\label{eq:inter cellular zak}
\end{eqnarray}
(See Appendix~\ref{app:zak phase} for further details). 
Note that $\gamma_n^{\mathrm{intra}}$ depends on the real-space origin while $\gamma_n^{\mathrm{inter}}$ does not, because $\alpha^{n,i,\zeta}_k$, being an element of the eigenvector of the momentum space representation of the Hamiltonian, is origin-independent.

To give a physical interpretation of the Zak phases just introduced, we start by relating the electronic part of the classical polarization of the unit cell to the \intracellular Zak phase.
We consider contributions from each filled band separately.
For the $n$-the band, we have
\begin{eqnarray}
P^{\text{el}}_{n,\mathrm{cl}} &=& \frac{1}{a}\int_\Omega dx~x \rho^{\text{el}}_{\mathrm{bulk},n}(x) \\
&=& -\frac{e}{a}\int_\Omega dx~x \sum_{k \in \mathrm{BZ}} \left| \psi_{n,k}(x)\right|^2  \\
&=& \frac{-e}{2\pi}\gamma_n^{\mathrm{intra}} - m_\Omega e.\label{eq:gamma_1_gen}
\end{eqnarray}
where $\rho^{\mathrm{el}}_{\mathrm{bulk},n}(x)$ is the electronic density corresponding to the $n$-th band.
That is, $P^{\mathrm{el}}_{n,\mathrm{cl}}$ can be evaluated from the intra-cellular Zak phase up to mod $e$.

Secondly, we give a physical interpretation of the \intercellular Zak phase.
To this end, we employ the Wannier functions given by
\begin{eqnarray}
W_{n,m}(x) &=& \frac{1}{\sqrt{N}} \sum_{k \in \mathrm{BZ}} \psi_{n,k}(x) e^{-\im kma} \\
&=& \sum_{m^\prime}^N\sum_{i=1}^{N_\mathrm{b}}\sum_{\zeta=1}^{N_{\mathrm{orb}}^i} A^{n,i,\zeta}_{m^\prime-m} \phi^{i,\zeta}_{m^\prime}(x)
\end{eqnarray}
where
\begin{eqnarray}
A^{n,i,\zeta}_{m^\prime-m} = \frac{1}{N}\sum_{k\in \mathrm{BZ}} e^{\im k(m^\prime-m)a} \alpha^{n,i,\zeta}_k .\label{eq:wannier_amp}
\end{eqnarray}
They satisfy the orthonormality condition $\langle W_{n,m}|W_{n^\prime,m^\prime}\rangle = \delta_{n,n^\prime}\delta_{m,m^\prime}$.
Note that in 1D the Wannier function $W_{n,m}(x)$ is in general guaranteed to be exponentially localized around a position in the $m$-th unit cell\cite{Kohn1959,Nenciu1983,Panati2007} while, in 2D and 3D, this is true if and only if the Chern number of the band vanishes.\cite{Kohmoto1985,Brouder2007}
Then, we express the $n$-th band \intercellular Zak phase, via the inverse transformation of~\eqref{eq:wannier_amp}, as
\begin{eqnarray}
\gamma_n^{\mathrm{inter}} &=& 2\pi \sum_{m}^N\sum_{i=1}^{N_\mathrm{b}}\sum_{\zeta=1}^{N_{\mathrm{orb}}^i} m \left| A^{n,i,\zeta}_{m} \right|^2.
\end{eqnarray}
By fixing a point $x_b$ at the boundary between neighbouring unit cells, which we assume for concreteness to be between the $m=-1$ and $m=0$ unit cells, the system is split into the left ($L$) and right ($R$) regions as depicted in Fig.~\ref{fig:left-right}.
With these definition, the \intercellular Zak phase can be split into two parts as
\begin{eqnarray}
\gamma_n^{\mathrm{inter}} = -\gamma^{R\rightarrow L}_n +\gamma^{L\rightarrow R}_n,
\end{eqnarray}
where
\begin{eqnarray}
\gamma^{R\rightarrow L}_n &=& -2\pi \sum_{m=-\infty}^{-1}\sum_{i=1}^{N_\mathrm{b}}\sum_{\zeta=1}^{N_\mathrm{orb}^i} m \left| A^{n,i,\zeta}_{m} \right|^2 \\ 
&=& 2\pi \sum_{m^\prime=0}^\infty \int_{-\infty}^{x_b} dx \left| W_{n,m^\prime}(x)\right|^2 \label{eq:wannier_rl}
\end{eqnarray}
and
\begin{eqnarray}
\gamma^{L\rightarrow R}_n &=& 2\pi \sum_{m=0}^{\infty}\sum_{i=1}^{N_\mathrm{b}}\sum_{\zeta=1}^{N_\mathrm{orb}^i} m \left| A^{n,i,\zeta}_{m} \right|^2 \\
&=& 2\pi \sum_{m^\prime=-\infty}^{-1} \int^{\infty}_{x_b} dx \left| W_{n,m^\prime}(x)\right|^2. \label{eq:wannier_lr}
\end{eqnarray}
Further details are given in Appendix~\ref{app:zak phase}.
From (\ref{eq:wannier_rl}) and (\ref{eq:wannier_lr}), we note that $Q^{R\rightarrow L} = -e/(2\pi)\sum_n^Z\gamma^{R\rightarrow L}_n$ represents the charge weight in the $L$ region of all Wannier functions with their centers belonging to the $R$ region, and vice versa for $Q^{L\rightarrow R} = -e/(2\pi)\sum_n^Z\gamma^{L\rightarrow R}_n$.
Finally, their difference in the $L$ region is simply expressed by
\begin{eqnarray}
\Delta Q^L = Q^{R\rightarrow L}-Q^{L\rightarrow R} =\frac{e}{2\pi} \sum_{n=1}^Z \gamma_n^{\mathrm{inter}}
\end{eqnarray}
where the sum is over the filled bands.
In the $R$ region, this difference is just given by $\Delta Q^R = -(e/2\pi) \sum_{n=1}^Z \gamma_n^{\mathrm{inter}}$.
Due to the translational symmetry of the system, this result is valid for any unit-cell boundary.
Also, this quantity does not depend on the position of the origin since the inter-cellular Zak phase does not as mentioned previously.
Note that, from the point of view of the electronic density, every unit cell is neutral and there is no extra charge in the $L$ and $R$ regions of the bulk.
However, we want to eventually obtain the extra charge accumulation around the surfaces when the system is terminated; we demonstrate how it is related to $Q^{R\rightarrow L}$ and $Q^{L\rightarrow R}$ in the next section.

%%%%%%%%%%%%%%%%%%%%%%%%%%%%%%%%%%%%%%%%%%%%%%%%%%%%%%%%%%
\begin{figure*}
\includegraphics[width=1.7\columnwidth]{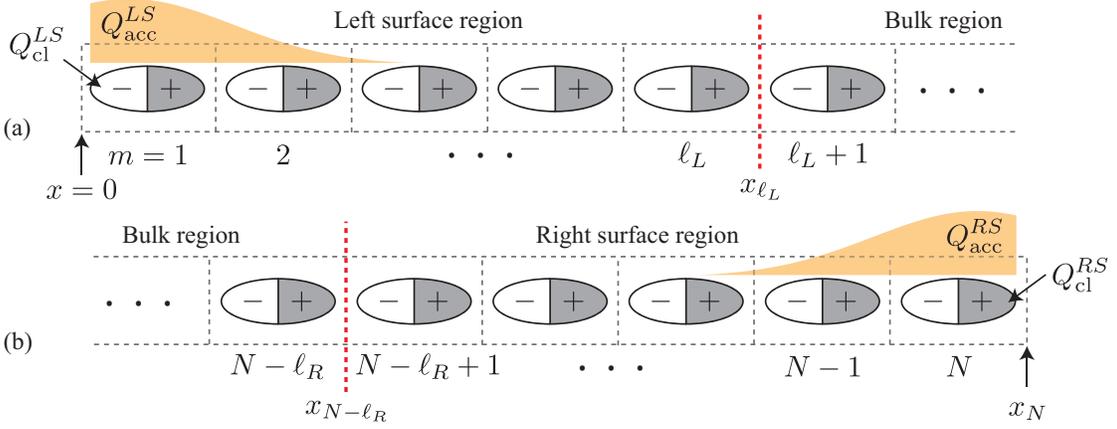}
\caption{(Color online) A schematic of the three regions (bulk, left and right surface region) of a finite 1D system with $N$ unit cells. Classical bound surface charges ($Q^{LS(RS)}_{\mathrm{cl}} = \mathbf{P}_{\mathrm{cl}}\cdot\hat{n}$) and extra charge accumulations ($Q^{LS(RS)}_{\mathrm{acc}}$) in two surface regions are also illustrated schematically. The system is terminated at $x=0$ and $x_N$. Boundaries between those three regions are represented by red dashed lines at $x_{\ell_L}$ and $x_{N-\ell_R}$. Bulk unit cells are drawn by gray dashed boxes and their indices are given below them. In each bulk unit cell, the  dipole moment ($\mathbf{P}_{\mathrm{cl}}$) of the bulk unit cell is drawn symbolically.}
\label{fig:finite_system}
\end{figure*}
%%%%%%%%%%%%%%%%%%%%%%%%%%%%%%%%%%%%%%%%%%%%%%%%%%%%%%%%%%

\section{Extra charge accumulations around surfaces}
\label{sec:extra_charge}

In this section we consider a terminated system, divided it into three regions as illustrated in Fig.~\ref{fig:finite_system}: the left surface region ($LS$) from $m=1$ to $m=\ell_L$ and the right surface ($RS$) region from $m=N-\ell_R+1$ to $m=N$, separated by the bulk region ($B$) from $m=\ell_L+1$ to $m=N-\ell_R$.
We choose $\ell_L$, $\ell_R$ and $N$ all large enough that there are no boundary effects in the bulk, and all regions are commensurate with the bulk's unit cell, namely they do not contain any partial unit cells.
While the extra charge accumulation in the surface region is defined as the additional charge over the bulk's charge distribution, it is equivalent to the total charge including ions in this region, because of the commensurability condition and the neutrality of the bulk's unit cell.
Finally, we assume that the finite system is also insulating because otherwise partially filled degenerate states at the Fermi level would yield ambiguity in evaluating physical quantities depending on which states we choose to be occupied.
Then, we show that, if the surface region is commensurate with the bulk's unit cell, the extra charge accumulation $Q^{LS(RS)}_\mathrm{acc}$ in the left (right) surface region of a neutral 1D insulator is given by
\begin{align}
Q^{LS(RS)}_\mathrm{acc} = +(-)\frac{e}{2\pi} \gamma^{\mathrm{inter}} ~~(\text{mod}~e) \label{eq:extra_charge_acc_inter_zak}
\end{align}
where $\gamma^{\mathrm{inter}}=\sum_{n=1}^Z \gamma_n^{\mathrm{inter}}$.

The overall strategy for the demonstration of the relation (\ref{eq:extra_charge_acc_inter_zak}) is as follows.
We prepare a set of orthonormal wave functions that are related to the set of occupied eigenfunctions of the finite system by a unitary transformation, following Vanderbilt et al.\cite{Vanderbilt1993b}
All of these states are localized, i.e., Wannier-like, such that their characteristic widths are much smaller than sizes of the three regions.
First, there are exchanges of weights of the Wannier functions between the surface and bulk region across their boundary as expressed by $\Delta Q^L$ and $\Delta Q^R$ in the previous section.
Then, from the orthonormality and localized feature of the prepared basis set, we show that the total charge in the left and right surface region should be $\Delta Q^L$ and $\Delta Q^R$ modulo $e$.

The orthonormal localized wave functions that we employ for the description of the terminated system are given by
\begin{align}
\boldsymbol\varphi = \big\{ &\varphi_1^{LS}(x),\cdots,\varphi_{s_L}^{LS}(x),W_{1,m_L}(x),\cdots,W_{Z,m_R}(x),\nonumber \\ &\varphi_1^{RS}(x),\cdots,\varphi_{s_R}^{RS}(x) \big\}^\mathrm{T}. \label{eq:localized_wavefunctions}
\end{align}
While the Wannier functions are the proper basis set for the infinite system, after the termination we need to construct this new set of basis wave functions that satisfy the open boundary condition.
$\boldsymbol\varphi$ span the same Hilbert space as the $M$ occupied eigenstates of the terminated system denoted by $\boldsymbol\psi^{\text{t.s.}}(x) = \left\{ \psi_1^{\text{t.s.}}(x),\cdots,\psi_M^{\text{t.s.}}(x) \right\}^\mathrm{T}$.\cite{Vanderbilt1993b}
First, we pick all occupied Wannier functions from the unit cell $m_L$ (in the left surface region) to $m_R$ (in the right surface region).
The choice of $m_L$ and $m_R$ is arbitrary but chosen to be far enough from the surfaces so that corresponding Wannier functions satisfy the boundary conditions with exponentially vanishing error.
We then replace the Wannier functions around the surfaces with another set of wave functions that satisfy the open boundary condition; these we denote by $\varphi_s^{LS}(x)$ and $\varphi_s^{RS}(x)$ around the left and right boundary each.
The set of wave functions in (\ref{eq:localized_wavefunctions}) can always be made orthonormal through the Gram-Schmidt orthonormalization process.\cite{Kohn1973,Rehr1974,Kallin1984,Vanderbilt1993b}

The charge density of the occupied electrons in the terminated system is given by $\rho^{\text{el}}_{\text{t.s.}}(x) = -e\boldsymbol\psi^{\text{t.s.}}(x)^\dag \boldsymbol\psi^{\text{t.s.}}(x)$.
Since $\boldsymbol\psi^{\text{t.s.}}(x)$ and $\boldsymbol\varphi(x)$, as orthonormal bases, span the same Hilbert space of occupied states, they are related by a unitary matrix, and the electronic charge is rewritten as
\begin{eqnarray}
\rho^{\text{el}}_{\text{t.s.}}(x) &=& -e\boldsymbol\varphi(x)^\dag \boldsymbol\varphi(x) \\
&=& -e\bigg( \sum_s^{s_L}|\varphi_s^{LS}(x)|^2 + \sum_s^{s_R}|\varphi_s^{RS}(x)|^2 \nonumber\\
&& + \sum_{n=1}^Z\sum_{m=m_L}^{m_R} |W_{n,m}(x)|^2\bigg).\label{eq:electronic density}
\end{eqnarray}
The total charge in the left surface region, which is the extra charge accumulation, is then evaluated as
\begin{eqnarray}
Q^{LS}_{\mathrm{acc}} &=&  \int_{0}^{x_{\ell_L}}dx \rho^{\text{el}}_{\text{t.s.}}(x) + \ell_{L}Ze
\end{eqnarray}
where $\ell_{L}Ze$ is the total ionic charge in the left surface region---we assume that the left-end of the system is at $x=0$ and $x_m = am$ is the position of the right boundary of the $m$-th unit cell.
In this region, integrals of $\varphi^{RS}_s(x)$'s in (\ref{eq:electronic density}) vanish, while the sum of integrals of $\varphi^{LS}_s(x)$'s, which are well localized to the right and left surface, is an integer number $s_L$.
Then, $Q^{LS}_{\mathrm{acc}}$ becomes
\begin{align}
Q^{LS}_{\mathrm{acc}} =& -e \int_{0}^{x_{\ell_L}}dx\sum_{n=1}^Z\sum_{m=m_L}^{m_R} |W_{n,m}(x)|^2 + q_Le \\
=& -e\sum_{n=1}^Z \sum_{m=m_L}^{\ell_L}\int_{0}^{x_{\ell_L}}dx|W_{n,m}(x)|^2 \nonumber\\
& -e\sum_{n=1}^Z \sum_{m=\ell_L+1}^{m_R}\int_{0}^{x_{\ell_L}}dx|W_{n,m}(x)|^2 +q_Le \label{eq:extra charge der1}
\end{align}
where $q_L=-s_L+\ell_L Z$.
Since all the Wannier functions are normalized, we have the identity
\begin{align}
&\sum_{n=1}^Z \sum_{m=m_L}^{\ell_L}\int_{0}^{x_{\ell_L}}dx|W_{n,m}(x)|^2 \nonumber\\
=&Z(\ell_L-m_L+1) - \sum_{n=1}^Z \sum_{m=m_L}^{\ell_L}\int^{x_{N}}_{x_{\ell_L}}dx|W_{n,m}(x)|^2. \label{eq:extra charge der2}\nonumber\\
\end{align}
This leads to the conclusion
\begin{align}
Q^{LS}_{\mathrm{acc}} =& \frac{-e}{2\pi}\sum_{n=1}^Z\left( \gamma^{R\rightarrow L}_n-\gamma^{L\rightarrow R}_n \right) + q^\prime_L e \\
=& \frac{e}{2\pi} \sum_{n=1}^Z \gamma_n^{\mathrm{inter}} + q^\prime_L e
\end{align}
where $q^\prime_L = q_L -Z(\ell_L-m_L+1)=-s_L+Z(m_L-1)$.
When we apply (\ref{eq:wannier_rl}) and (\ref{eq:wannier_lr}) to (\ref{eq:extra charge der1}) and (\ref{eq:extra charge der2}), we take $m_L\rightarrow -\infty$, $m_R\rightarrow \infty$, $x_1\rightarrow -\infty$, and $x_{N}\rightarrow \infty$, which is an approximation justified by the fact that the Wannier functions are exponentially localized in space and that the surface regions can be taken large enough. 
Also, the role of $x_b$ in (\ref{eq:wannier_rl}) and (\ref{eq:wannier_lr}) is taken by $x_{\ell_L}$ in (\ref{eq:extra charge der1}) and (\ref{eq:extra charge der2}).
In the same way, the extra charge accumulation in the right surface region is given by
\begin{eqnarray}
Q^{RS}_{\mathrm{acc}} = -\frac{e}{2\pi} \sum_{n=1}^Z \gamma_n^{\mathrm{inter}} + q^\prime_R e
\end{eqnarray}
where $q^\prime_R = -s_R + Z(N-m_R)$.
As in Ref.~\onlinecite{Vanderbilt1993b}, we are not interested in the integer numbers $q^\prime_L$ and $q^\prime_R$ which cannot be determined from the bulk.
Instead, we conclude that the inter-cellular Zak phase can predict the extra charge accumulation in the surface region modulo $e$ as described in (\ref{eq:extra_charge_acc_inter_zak}).

Finally, we remark on the difference between the bound surface charge and the extra charge accumulation.
The bound surface charge $\sigma$ is a quantity measured in a capacitance measurement and is related to the modern definition of the polarization by $\sigma = \mathbf{P}\cdot\hat{n}$, where $\hat{n}$ is the surface orientation ($\hat{n} = -(+)\hat{x}$ for the left (right) edge in the 1D case).\cite{Vanderbilt1993b}
The bound surface charge of the total charge density $\rho_{\text{t.s.}}(x)$ of the finite system is evaluated explicitly as\cite{Vanderbilt1993b,Resta1988,Resta2007b}
\begin{eqnarray}
\sigma^{LS} = \frac{1}{a}\int_{-\infty}^{x_c}dx \int_{x - \frac{a}{2}}^{x + \frac{a}{2}}dx^\prime \rho_{\text{t.s.}}(x^\prime)\label{eq:left bound charge 1}
\end{eqnarray}
for the bound surface charge at the left edge, and
\begin{eqnarray}
\sigma^{RS} = \frac{1}{a}\int^{\infty}_{x_d}dx \int_{x - \frac{a}{2}}^{x + \frac{a}{2}}dx^\prime \rho_{\text{t.s.}}(x^\prime)\label{eq:left bound charge 2}
\end{eqnarray}
at the right edge, where $x_c$ and $x_d$ are arbitrary positions in the middle of the finite system far away from the surfaces.
As shown in Appendix \ref{app:bound charge}, $\sigma^{LS}$ and $\sigma^{RS}$ are independent of $x_c$ and $x_d$.
To clearly compare the bound surface charge and the extra charge accumulation, we choose them as $x_c=x_{\ell_L}+a/2$ and 
$x_d=x_{N-\ell_R}-a/2$.
Then the left and right bound surface charge becomes
\begin{align}
\sigma^{LS} = -\frac{1}{a}\int^{x_{\ell_L}+a}_{x_{\ell_L}}dx~x\rho_{\text{t.s.}}(x) + \int_{0}^{x_{\ell_L}}dx\rho_{\text{t.s.}}(x),\label{eq:left bound charge}
\end{align}
and
\begin{align}
\sigma^{RS} = \frac{1}{a}\int^{x_{N-\ell_R}}_{x_{N-\ell_R}-a}dx~x\rho_{\text{t.s.}}(x) + \int^{x_{N}}_{x_{N-\ell_R}}dx\rho_{\text{t.s.}}(x).\label{eq:right bound charge}
\end{align}
Derivations of the above are given in Appendix \ref{app:bound charge}.
The first terms of (\ref{eq:left bound charge}) and (\ref{eq:right bound charge}) are exactly $\mathbf{P}_{\mathrm{cl}}\cdot\hat{n}$ because $x_{\ell_L}$ and $x_{N-\ell_R}$ are far enough from the edges so that $\rho_{\text{t.s.}}(x)$ can be considered a bulk charge density.
We call them the \textit{classical bound surface charges} $Q^{LS(RS)}_\mathrm{cl}$ because in classical electrodynamics they are the bound surface charges in a dielectric material with a uniform dipole distribution through the whole finite system.
On the other hand, the second terms of (\ref{eq:left bound charge}) and (\ref{eq:right bound charge}) are just the total charge in the left and right surface regions, which are equal to the extra charge accumulations in those regions in the neutral systems.
Consequently, we have shown that the bound surface charge consists of two contributions, one from the bulk's dipole moment, and the other from the extra charge accumulation.
This is consistent with the splitting of the Zak phase into the intra- and inter-cellular Zak phase.
Also, it indicates that the modern polarization is actually composed of two kinds of polarizations, one is the classical one from the bulk's dipole moment, and the other is polarization from the extra charge accumulations at opposite edges.
This conclusion is illustrated schematically in Fig.~\ref{fig:finite_system}.

\section{Bulk-boundary correspondence}
\label{sec:bulk_boundary}

Our general statement about the number of surface modes in the gap of a finite 1D insulator with charge neutrality is as follows:
there are $n_s = \gamma^{\mathrm{inter}}/\pi$ (mod 2) surface modes below the Fermi level if there is inversion symmetry both before and after termination, and the finite system is commensurate with the bulk unit cell used for the calculation of $\gamma^{\mathrm{inter}}$.
We justify our general statement in the light of the physical interpretation of the inter-cellular Zak phase $\gamma^{\mathrm{inter}}$.
As discussed in the previous section, the inter-cellular Zak phase explains the amount of the extra charge accumulation around the edge which is closely related to the surface modes, while the intra-cellular Zak phase, as a bulk dipole moment, has nothing to do with the surface modes.

First, we note that if the bulk respects inversion symmetry, $\gamma^\mathrm{inter}$ is quantized to $\pi$.
When the real-space origin is at one of the inversion centers, the total Zak phase $\gamma$ is quantized to $\pi$ while its intra-cellular part vanishes because the dipole moment of the bulk unit cell is zero in this case.
Therefore, $\gamma^\mathrm{inter} = \gamma$ is also quantized to $\pi$, and it is independent of the choice of the origin as mentioned in Sec.~\ref{sec:reinterpretation}.

The surface modes are the eigenstates generated in the bulk gaps as a result of the edge termination.
As described in the previous section, they are exponentially localized at one of the edges.
Due to inversion symmetry, if we have a surface state localized at the left edge, we always find its counterpart localized at the right edge with the same energy.
The degeneracy might increase if the system preserves additional symmetries such as the time reversal symmetry.

We proceed to the justification of our general statement by examining two different cases.
One is when the Fermi level of the finite system is not at any of the surface modes, so that the finite system is also an insulator.
And, the other is the case when the Fermi level is at the partially filled degenerate surface modes.

First, let us consider the case when the finite system is already insulating.
Then, the only allowed value of the total charge accumulations is $Q_\mathrm{acc}^{LS}=Q_\mathrm{acc}^{RS}=0$ due to charge neutrality and inversion symmetry.
According to (\ref{eq:extra_charge_acc_inter_zak}), the \intercellular Zak phase should be 0 (mod $2\pi$) in this case.
Since every surface mode has an inversion partner with the same energy, there should be an even number of surface states below the Fermi level.
This means the general statement holds for insulating phases with inversion symmetry.

Second, we consider the case when the Fermi level is located at the surface modes.
In this case, we cannot use the results of the previous section since there we assume the finite system is insulating.
However, we can resolve this obstacle by opening tiny gaps between degenerate surface modes without changing the total number of surface modes in the bulk gap.
We assume that this can be done by applying local perturbative potentials on both edges that break symmetries corresponding to those degeneracies. 
For instance, different on-site potentials at opposite edges would break the degeneracy responsible for inversion symmetry.
Similarly, a local Zeeman field could be used to break spin degeneracy.

Let us investigate the dependence of the surface modes on the tiny inversion symmetry breaking, which is relevant for the quantization of the extra charge accumulations on the edges as explained below.
This is achieved by applying an on-site potential $\delta$ to the first unit cell and $-\delta$ to the  $N$-th unit cell of the finite system where $\delta$ is nonzero but small compared with the bulk gap.
Then, the eigenstates at opposite edges have different energies, and it is impossible to construct surface modes localized to both edges simultaneously from them.  
As a result, the eigenstate of each lifted surface mode is localized to only one of the left and right edges.
This implies that each surface mode yields and integer charge $Q_{S_i,\text{acc}}^{LS}=-e$ or $Q_{S_i,\text{acc}}^{RS}=-e$, where $Q_{S_i,\text{acc}}^{LS(RS)}$ is the contribution of the $i$-th surface mode to the extra charge accumulation in the left (right) surface region.
After taking into account the inversion symmetry breaking, we assume that other perturbative potentials would not affect this property of the lifted surface modes.

On the other hand, the charge distribution calculated from the bulk band continuum is insensitive against this kind of local perturbation due to its bulk character.
So, the charge distribution of the bulk band continuum maintains almost the same form as the unperturbed inversion symmetric system.
In the unperturbed inversion symmetric system, the extra charge accumulation from the bulk band continuum can only take values of $Q_{B,\text{acc}}^{LS}=Q_{B,\text{acc}}^{RS}=(p+1/2)e$ or $pe$ where $p$ is an integer, because the total charge of the finite system should be an integer multiple of $e$ and the system has inversion symmetry.
Here, $Q_{B,\text{acc}}^{LS(RS)}$ is the contribution of the bulk band continuum to the extra charge accumulation in the left (right) surface region.
Note that although we slightly break the inversion symmetry to make the system insulating, its presence in the unperturbed system is essential for the constraint on $Q_{B,\text{acc}}^{LS(RS)}$.

If $Q_{B,\text{acc}}^{LS(RS)}=(p+1/2)e$, the total charge accumulations originating from surface modes on both sides should be $\sum_{E_i<E_F}(Q_{S_i,\text{acc}}^{LS} + Q_{S_i,\text{acc}}^{RS}) = -(2p+1)e$ to maintain neutrality.
This means that there is an odd number of surface modes below the Fermi level in the main gap where the Fermi level lies because there were even number of occupied surface modes in other gaps before breaking inversion symmetry, and we have assumed that this number is unchanged by the symmetry breaking perturbations.
Also, due to the same reason, this implies that we have an odd number of surface modes below the Fermi energy in the main gap before breaking those symmetries.
In the perspective of the inter-cellular Zak phase, the fact that the extra charge accumulation $Q_{B,\text{acc}}^{LS(RS)} + \sum_{E_i<E_F}Q_{S_i,\text{acc}}^{LS(RS)}$ is a half-integer multiple of $e$ implies that the inter-cellular Zak phase should be a half-integer multiple of $2\pi$ according to (\ref{eq:extra_charge_acc_inter_zak}).
In conclusion, the number of surface modes is equal to $\gamma^{\mathrm{inter}}/\pi$  mod 2 in this case.

In similar fashion, when $Q_{B,\text{acc}}^{LS(RS)}=pe$, the surface modes' extra charge accumulations on both edges should be $\sum_{E_i<E_F}(Q_{S_i,\text{acc}}^{LS} + Q_{S_i,\text{acc}}^{RS}) = -2pe$  to maintain charge neutrality, implying that there is an even number of surface modes in the main gap and that the inter-cellular Zak phase is even.
So, the number of surface modes is again equal to $\gamma^{\mathrm{inter}}/\pi$ mod 2 in this case.

Finally, we note that this bulk-boundary correspondence can be restated by using the Zak phase when the real-space origin is at an inversion center.
This is because the intra-cellular Zak phase becomes zero (mod $2\pi$) according to (\ref{eq:intra cellular zak}), so that the Zak phase is identical to the inter-cellular Zak phase in this case.
%

%%%%%%%%%%%%%%%%%%%%%%%%%%%%%%%%%%%%%%%%%%%%%%%%%%%%%%%%%%
\begin{figure}
\includegraphics[width=1\columnwidth]{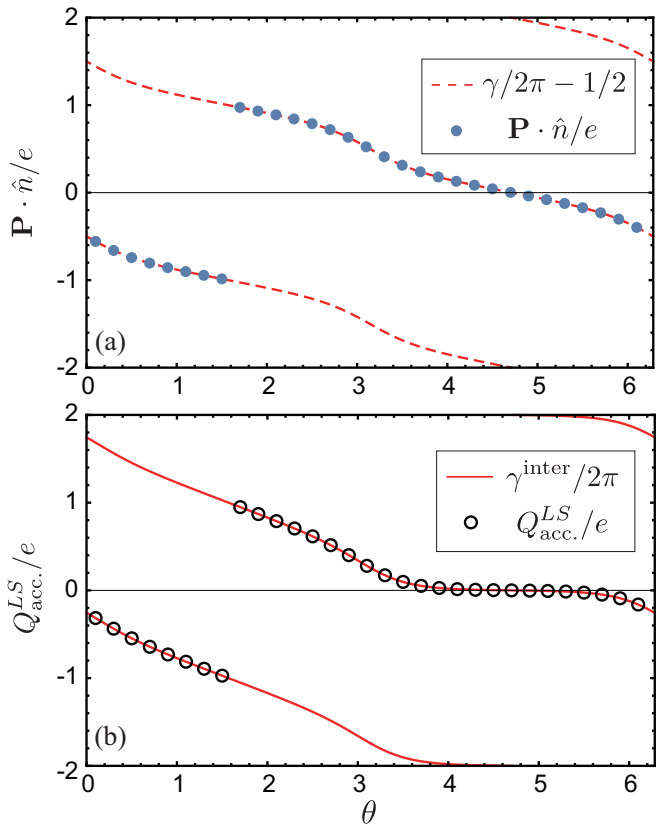}
\caption{(Color online) The comparison between the Zak phase, the \intercellular Zak phase, the bound surface charge and the extra charge accumulation of the Rice-Mele 1D chain model. Here, the tight-binding parameters are given by $t=1$, $\Delta=0.6\cos\theta$ and $\delta=0.6\sin\theta$. (a) The red dashed curves are the calculations of $\gamma/2\pi - 1/2$ (mod $2e$). The blue dots are the bound surface charge at the left edge of the terminated chain. (b) The red solid curves are the inter-cellular Zak phase and the circular markers are the extra charge accumulation near the left edge.}
\label{fig:rice_mele}
\end{figure}
%%%%%%%%%%%%%%%%%%%%%%%%%%%%%%%%%%%%%%%%%%%%%%%%%%%%%%%%%%
\section{Examples}
\label{sec:examples}
In this section we provide a few examples demonstrating the general results of the earlier sections.

\subsection{Rice-Mele 1D chain}
\label{subsec:rice-mele}

In this subsection, we calculate the extra charge accumulation in the Rice-Mele model\cite{Vanderbilt1993b,Rice1982} and show that it is accurately predicted by the inter-cellular Zak phase and is different from the bound surface charge.
Also, for inversion symmetric cases of this model, we show that the bulk-boundary correspondence works well.

The Rice-Mele model is given by
\begin{eqnarray}
\mathcal{H}_{\mathrm{RM}} = \sum_{\sigma,j} \epsilon_j c^\dag_{\sigma j} c_{\sigma j} + \sum_{\sigma,j} \left[ V_{j,j+1}c^\dag_{\sigma j} c_{\sigma j+1} + \mathrm{h.c.}\right]~~
\end{eqnarray}
where $\sigma$ and $j$ are indices for the spin and the lattice sites.
For neutrality of the system at half-filling, we introduce spin degeneracy.
The tight binding parameters are given by $\epsilon_{2p}=-\Delta$, $\epsilon_{2p+1}=\Delta$, $V_{2p-1,2p}=-t-\delta$ and $V_{2p,2p+1}=-t+\delta$ where $p$ is an integer.
Note that there are two sites per unit cell.
When $\Delta =0$, the model reduces to the Su-Schrieffer-Heeger model\cite{Su1979}.

The Bloch wave function for $\mathcal{H}_{\mathrm{RM}}$ is described by
\begin{eqnarray}
\psi_{n,k}(x) = \frac{1}{\sqrt{N}} \sum_m^N e^{ikm} \left[ \alpha_{n,k}\phi^A_m(x) + \beta_{n,k}\phi^B_m(x)\right]~~ \label{eq:bloch_RM}
\end{eqnarray}
where
\begin{eqnarray}
\phi^A_m(x) = \phi(x-m) ~\mathrm{and}~ \phi^B_m(x) = \phi(x-m-\frac{1}{2})
\end{eqnarray}
and $\phi(x)$ is the atomic wave function at the lattice sites.
In this section we set $a=1$ for convenience.

The intra- and inter-cellular Zak phases are given by
\begin{eqnarray}
\gamma_{n,\sigma}^{\mathrm{intra}} = \int_0^{2\pi}dk \frac{1}{2}|\beta_{n,k}|^2
\end{eqnarray}
and
\begin{eqnarray}
\gamma_{n,\sigma}^{\mathrm{inter}} = \im \int_0^{2\pi}dk \left( \alpha^*_{n,k}\frac{\partial \alpha_{n,k}}{\partial k} + \beta^*_{n,k}\frac{\partial \beta_{n,k}}{\partial k} \right). \label{eq:zak_phase1}
\end{eqnarray}
Since there is no couplings between the two spin species, we have $\gamma_{n,\uparrow} = \gamma_{n,\downarrow}$.
According to the results in Sec.~\ref{sec:reinterpretation}, the classical polarization for a given unit cell is
\begin{eqnarray}
P_\text{cl} = \frac{-e}{2\pi}\sum_{\sigma}\sum_{n=1}^Z \gamma_{n,\sigma}^{\mathrm{intra}} + \frac{e}{2}\label{eq:rice_mele_classical_polarization}
\end{eqnarray}
and the extra charge accumulations at the edges are
\begin{eqnarray}
Q^{LS(RS)}_{\text{acc}} = +(-)\frac{e}{2\pi}\sum_{\sigma}\sum_{n=1}^Z \gamma_{n,\sigma}^{\mathrm{inter}}~~(\mathrm{mod} ~2e).
\end{eqnarray}
Here, $Z=1$ since we consider two spins independently.

We check this relation between $Q^{LS(RS)}_{\text{acc}}$ and the \intercellular Zak phase numerically in Fig.~\ref{fig:rice_mele}(b) where the system is parameterized by $t=1$, $\Delta=0.6\cos\theta$ and $\delta=0.6\sin\theta$.
Also, for the bound surface charge ($\sigma = \mathbf{P}\cdot\hat{n}$), we applied the formula\cite{Resta1988,Resta2007b} (\ref{eq:left bound charge 1}) and (\ref{eq:left bound charge 2}) and reproduced exactly the results obtained by Vanderbilt et al.\cite{Vanderbilt1993b} as shown in Fig.~\ref{fig:rice_mele}(a).
The difference between those two quantities corresponds to the classical bound surface charge $\sigma_\text{cl} = \mathbf{P}_\text{cl}\cdot\hat{n}$ originating from the dipole moment of the unit cell as we discussed in Sec.~\ref{sec:extra_charge}.
Since the dipole moment vanishes when the unit cell respects inversion symmetry, those two quantities become identical to each other at $\theta = \pi/2$ and $\theta = 3\pi/2$ where $\Delta = 0$ as shown in Fig.~\ref{fig:rice_mele}.

As in Fig.~\ref{fig:rice_mele}(b), the \intercellular Zak phase becomes $2\pi$ for $\theta=\pi/2$ and zero for $\theta=3\pi/2$, which predicts an even number of surface bands below the Fermi level according to the bulk-boundary correspondence in the previous section.
In the finite system with an even number of sites, we find two surface bands for $\theta=\pi/2$ and no surface bands for $\theta=3\pi/2$ below the Fermi energy (data not shown), consistent with the bulk-boundary correspondence.
However, if we consider two spins separately, we have more information.
For each spin species, the \intercellular Zak phase is just the half of those in Fig.~\ref{fig:rice_mele}(b), namely $\pi$ for $\theta=\pi/2$ and zero for $\theta=3\pi/2$ which implies there is an odd and even number of surface states in each case.
This means we must have surface modes when $\theta=\pi/2$ while one cannot determine the existence of the surface bands when $\theta=3\pi/2$.
%

%%%%%%%%%%%%%%%%%%%%%%%%%%%%%%%%%%%%%%%%%%%%%%%%%%%%%%%%%%
\begin{figure}
\includegraphics[width=1\columnwidth]{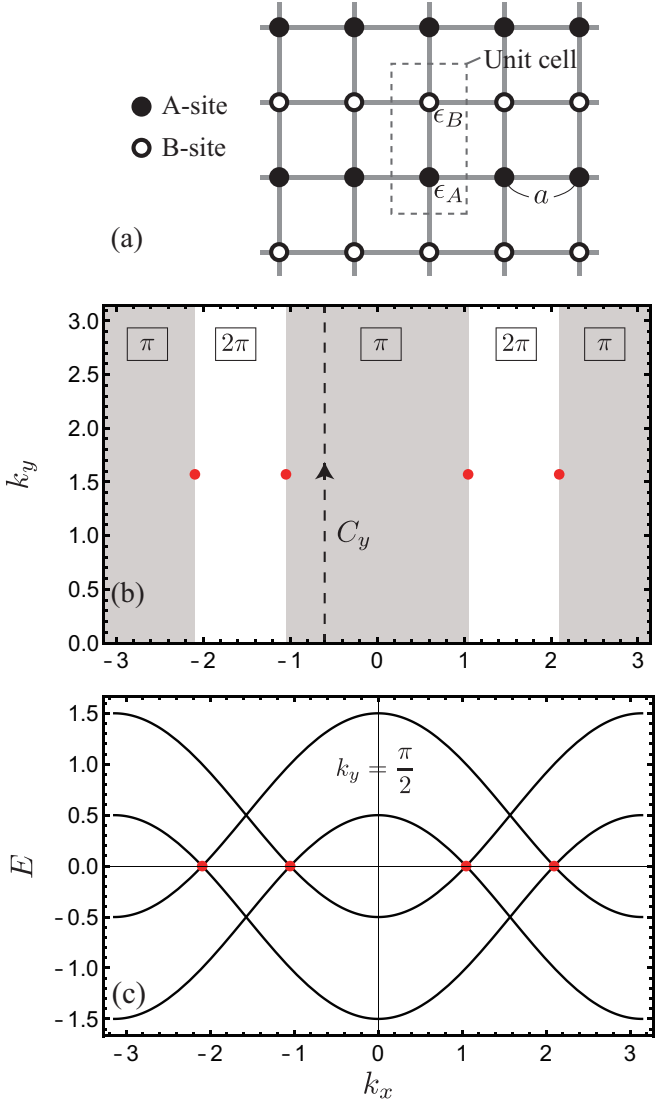}
\caption{(Color online)  The lattice structure of the 2D toy model-I. The unit cell used for the calculations of the \intercellular Zak phases is the dashed box. (b) The Brillouin zone of the given system. Four Dirac nodes at $k_y=\pi/2$ are marked by red dots. The Brillouin zone is divided into several regions depending on the value of the Zak phase: $\gamma = \pi$ is gray and $\gamma = 2\pi$ is white. (c) Band spectrum at $k_y=\pi/2$. The parameters used are $t=\epsilon_0 = 0.5$. 
}
\label{fig:toy2d_band}
\end{figure}
%%%%%%%%%%%%%%%%%%%%%%%%%%%%%%%%%%%%%%%%%%%%%%%%%%%%%%%%%%

%%%%%%%%%%%%%%%%%%%%%%%%%%%%%%%%%%%%%%%%%%%%%%%%%%%%%%%%%%
\begin{figure}
\includegraphics[width=1\columnwidth]{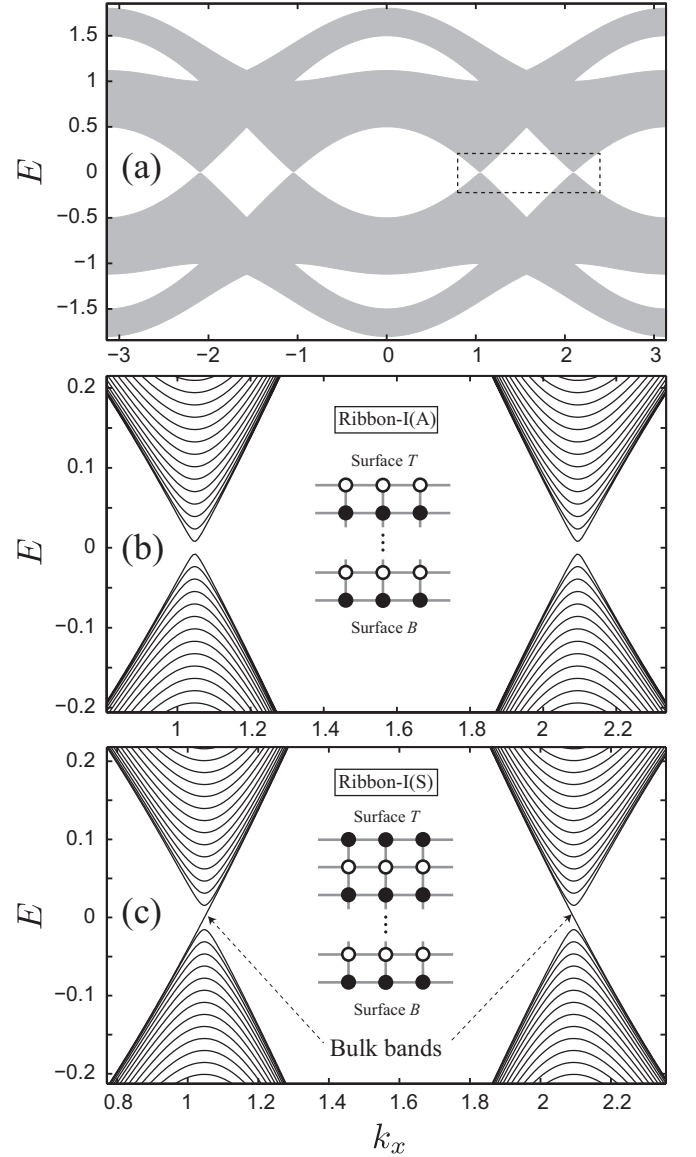}
\caption{(a) A schematic band dispersion of the terminated system of the 2D toy model-I. The dashed box in (a) is highlighted in (b) and (c) for different choices of the terminations, Ribbon-I(A) and Ribbon-I(S) each whose lattice structures are depicted in the insets of (b) and (c). Those ribbon geometries are infinite only along the $x$ axis. Their bottom and top surfaces are denoted by surface-$B$ and $T$.}
\label{fig:toy2d_band_surf}
\end{figure}
%%%%%%%%%%%%%%%%%%%%%%%%%%%%%%%%%%%%%%%%%%%%%%%%%%%%%%%%%%

\subsection{2D toy model-I}
\label{subsec:2d_toy}

Here, we suggest a 2D toy model as a counterexample for the conventional bulk-boundary correspondence of the Zak phase.
From this, we clarify its right usage by emphasizing the assumption of the existence of the commensurate bulk's unit cell.
We also test whether the inter-cellular Zak phase predicts the extra charge accumulation accurately or not.

Our toy model-I is illustrated in Fig.~\ref{fig:toy2d_band}(a) and described by the Hamiltonian
\begin{align}
\mathcal{H}_{\mathrm{2D}}^{\mathrm{I}} &= \sum_{\mathbf{R},i}\Big( \epsilon_A a^\dag_{i,\mathbf{R}}a_{i,\mathbf{R}} + \epsilon_B b^\dag_{i,\mathbf{R}}b_{i,\mathbf{R}} \Big) \nonumber \\ 
& + t\sum_{\mathbf{R},\boldsymbol\delta_x} \Big( a^\dag_{2,\mathbf{R}+\boldsymbol\delta_x} a_{1,\mathbf{R}} + b^\dag_{2,\mathbf{R}+\boldsymbol\delta_x} b_{1,\mathbf{R}} + \mathrm{h.c.} \Big) \\
& + t\sum_{\mathbf{R},\boldsymbol\delta_y} \Big( a^\dag_{1,\mathbf{R}+\boldsymbol\delta_y} b_{1,\mathbf{R}} - a^\dag_{2,\mathbf{R}+\boldsymbol\delta_y} b_{2,\mathbf{R}} + \mathrm{h.c.} \Big), \nonumber
\end{align}
where $\mathbf{R}$ is the position vector of the unit cell, and $\boldsymbol\delta_x = \pm a \hat{x}$ and $\boldsymbol\delta_y = \pm a \hat{y}$ are the nearest neighbor vectors.
There are two orbitals at each site denoted by $i=1,~2$
The electron hops between different orbitals along $x$ and between the same orbitals along $y$.
Along the $y$ direction, the sign of the hopping parameter of the $i=1$ and $i=2$ orbitals are opposite.

For $\epsilon_A = -\epsilon_B = \epsilon_0$, the energy spectrum is given by
\begin{eqnarray}
E_{\mathrm{2D}}^{\mathrm{I}}(k_x,k_y) &=& \pm\left\{ \left( \epsilon_0 \pm 2t\cos k_x \right)^2 + 4t^2\cos^2k_y\right\}^{\frac{1}{2}}~
\end{eqnarray}
where $a$ is set to be unity.
The spectrum can become gapless only at $k_y = \pi/2$, where it becomes $E_{\mathrm{2D}}^{\mathrm{I}}(k_x,\pi/2) = \pm(\epsilon_0 \pm 2t\cos k_x)$ which has four Dirac nodes at $k_x= \pm \cos^{-1}(\pm\epsilon_0/2t)$ for $0 < \epsilon_0 < 2t$ as shown in Fig. \ref{fig:toy2d_band}(b) and (c) where $t=\epsilon_0 = 0.5$.

For the Zak phase analysis, we consider an effective 1D Hamiltonian obtained by fixing the momentum $k_x$.
From this, we can calculate the Zak phase by integrating the Berry connection along $k_y$.
The results are presented in Fig. \ref{fig:toy2d_band}(b) by the gray ($\gamma = \pi$) and white ($\gamma = 2\pi$) regions.
Due to reflection symmetry, which corresponds to inversion symmetry in the effective Hamiltonian, we have quantized Zak phases.
Whenever we cross one of the Dirac nodes, the Zak phase changes discontinuously by $\pm\pi$.

%%%%%%%%%%%%%%%%%%%%%%%%%%%%%%%%%%%%%%%%%%%%%%%%%%%%%%%%%%
\begin{figure}
\includegraphics[width=1\columnwidth]{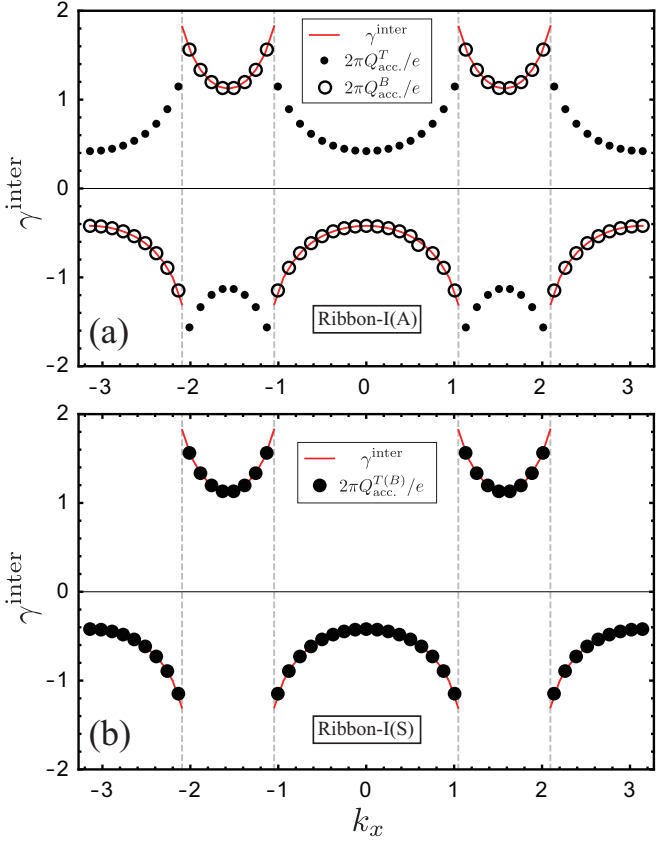}
\caption{(Color online) The \intercellular Zak phase ($\gamma^\text{inter} = \sum_n\gamma^\text{inter}_n$) is plotted by red solid curves as a function of $k_x$ and it is compared with the extra charge accumulation which is represented by empty and filled circular markers. In (a), the empty (filled) circles are $2\pi Q^B_{\text{acc}}/e$ ($2\pi Q^T_{\text{acc}}/e$) of Ribbon-I(A). In (b), the filled circles represent the densities of the extra charge accumulations on both sides of Ribbon-I(S). In this case, the extra charge accumulations on both surfaces is the same due to reflection symmetry.}
\label{fig:toy2d_zak}
\end{figure}
%%%%%%%%%%%%%%%%%%%%%%%%%%%%%%%%%%%%%%%%%%%%%%%%%%%%%%%%%%

%%%%%%%%%%%%%%%%%%%%%%%%%%%%%%%%%%%%%%%%%%%%%%%%%%%%%%%%%%
\begin{figure}
\includegraphics[width=1\columnwidth]{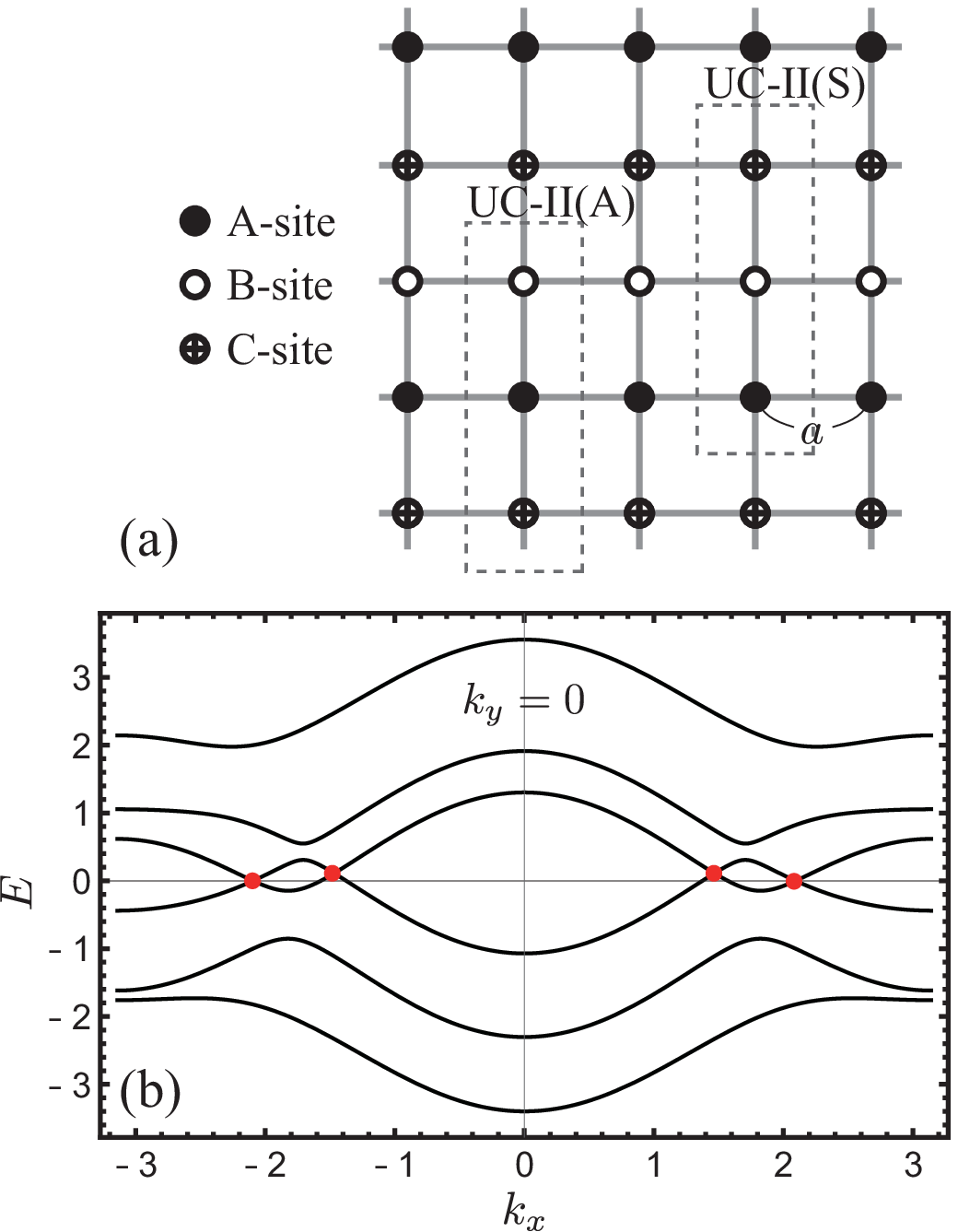}
\caption{(Color online) (a) The lattice structure of the 2D toy model-II. Two unit cell choices for the evaluation of the \intercellular Zak phase are shown by two dashed boxes. (b) The band structure at $k_y=0$ for $t=0.5$ and $H_0=\mathrm{diag}(0.5,-1.5,2,0,0.5,-1.5)$. Four Dirac points are represented by red dots.}
\label{fig:toy2d_3band}
\end{figure}
%%%%%%%%%%%%%%%%%%%%%%%%%%%%%%%%%%%%%%%%%%%%%%%%%%%%%%%%%%

Now, let us consider ribbon-shaped finite systems that are finite in the $y$-direction but infinite in the $x$-direction.
We consider two kinds of ribbon geometries, which we call Ribbon-I(A) and Ribbon-I(S) as illustrated in the insets of Fig.~\ref{fig:toy2d_band_surf}(b) and (c).
Ribbon-I(A) does not respect reflection symmetry along the $x$ axis while Ribbon-I(S) does.
As a result, only Ribbon-I(S) remains gapless without breaking the Dirac nodes of the bulk.

According to the conventional bulk-boundary correspondence, we expect Ribbon-I(S) to have surface modes at $k_x$'s in the gray or white region in Fig.~\ref{fig:toy2d_band}(b) because both the bulk and terminated system have reflection symmetry.
However, it turns out that there are no boundary modes in the Brillouin zone for both ribbon geometries as shown in Fig.~\ref{fig:toy2d_band_surf}.
This is a good example of the wrong usage of the bulk-boundary correspondence using the Zak phase.
In the revised correspondence given in the previous section, it requires the commensurability between the finite system and the bulk's unit cell in addition to inversion symmetry.
Ribbon-I(A) is commensurate with the unit cell described in Fig.~\ref{fig:toy2d_band}(a), but does not respect mirror symmetry while Ribbon-I(S), which is reflection symmetric, cannot be commensurate with any bulk's unit cell as it consists of an odd number of dimer lines.
Therefore, those two systems are actually beyond the scope of the revised bulk-boundary correspondence.

We check that the \intercellular Zak phase accurately predicts the extra charge accumulation in the surface regions in Fig. \ref{fig:toy2d_zak}.
From (\ref{eq:extra_charge_acc_inter_zak}), the extra charge accumulation for a given $k_x$ is given by
\begin{eqnarray}
Q^{B(T)}_{\text{acc}}(k_x) = \pm\frac{e}{2\pi}\sum_{n} \gamma_{n}^{\mathrm{inter}}(k_x)~~(\mathrm{mod} ~e)
\end{eqnarray}
where $B(T)$ and the plus(minus) sign correspond to the bottom(top) surface, and the sum is over occupied bands.
In the case of Ribbon-I(A), both bottom and top surfaces are commensurate with a single unit cell and share the same \intercellular Zak phase.
As a result, the extra charge accumulations in the bottom and top surfaces show opposite signs to each other as shown in Fig. \ref{fig:toy2d_zak}(a).
On the other hand, for Ribbon-I(S), the commensurate unit cell for the top surface is the one obtained by shifting the commensurate unit cell for the bottom by $a\hat{y}$.
We find that the \intercellular Zak phases for those unit cells have the same magnitude but opposite signs.
While the extra charge accumulations in the bottom and top surfaces should be identical due to the mirror symmetry, it is consistent with the above property of \intercellular Zak phases.
This is described in Fig. \ref{fig:toy2d_zak}(b).

\subsection{2D toy model-II}
\label{subsec:2d_toy_II}

%%%%%%%%%%%%%%%%%%%%%%%%%%%%%%%%%%%%%%%%%%%%%%%%%%%%%%%%%%
\begin{figure}
\includegraphics[width=1\columnwidth]{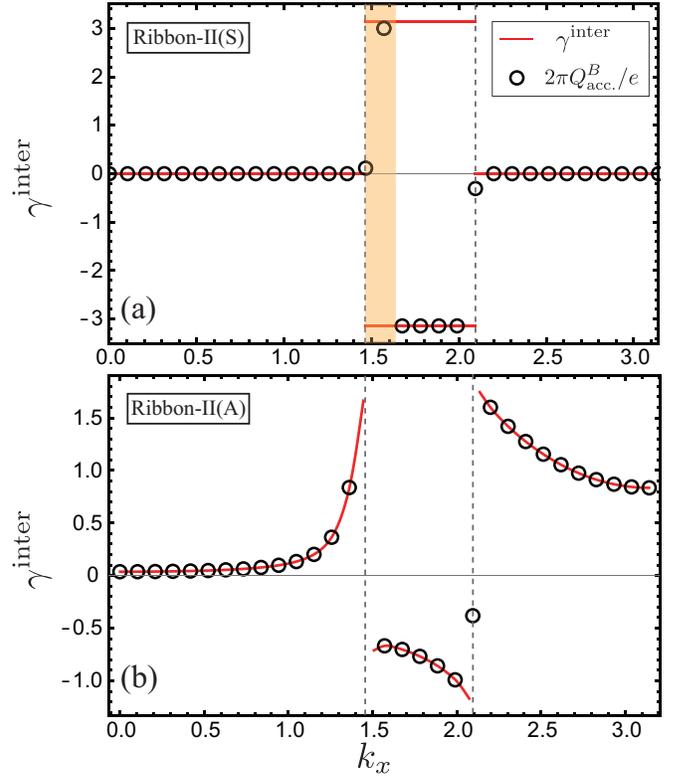}
\caption{(Color online) Calculations of the \intercellular Zak phases for (a) UC-II(S) and (b) UC-II(A). They are compared with the extra charge accumulations in the bottom surfaces of Ribbon-II(S) and Ribbon-II(A) which are marked by empty circles. The positions of the Dirac nodes are shown by vertical dashed lines. The shaded region in (a) is where the bulk-boundary correspondence fails because the Fermi level of the 1D effective Hamiltonian for given $k_x$ is located at the edge of the bulk continuum as shown in Fig. \ref{fig:toy2d_3band_surf}(a) and (b).}
\label{fig:toy2d_3zak}
\end{figure}
%%%%%%%%%%%%%%%%%%%%%%%%%%%%%%%%%%%%%%%%%%%%%%%%%%%%%%%%%%

%%%%%%%%%%%%%%%%%%%%%%%%%%%%%%%%%%%%%%%%%%%%%%%%%%%%%%%%%%
\begin{figure}
\includegraphics[width=1\columnwidth]{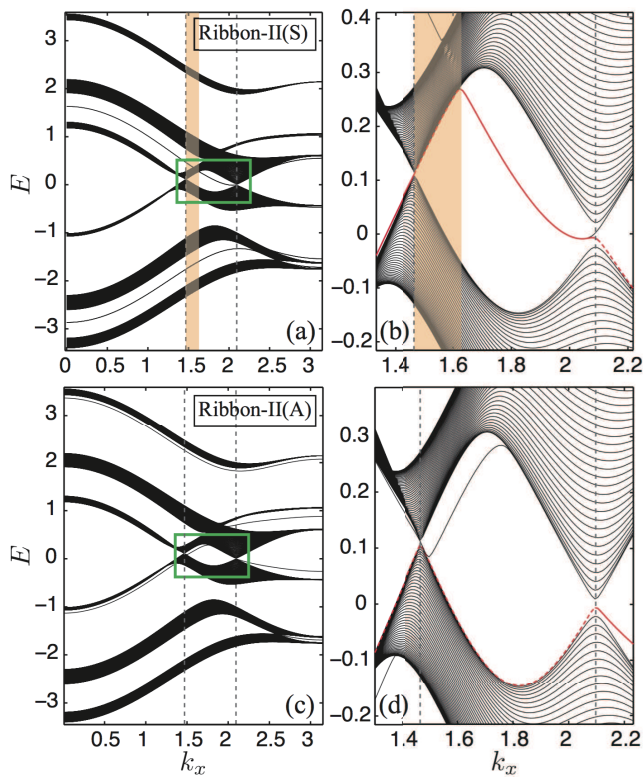}
\caption{(Color online) Energy spectra of Ribbon-II(S) and Ribbon-II(A) are plotted in (a) and (c). A zoom into the region covered by the green boxes is given in (b) and (c). The meaning of the vertical dashed lines and the shaded regions is explained in Fig. \ref{fig:toy2d_3zak}. In (b) and (d), the red curves are the topmost filled band of the effective 1D system for given $k_x$ at half-filling. If the wave function of this band is localized at the surface, we represent it by a solid curve. On the other hand, if the eigenstate is bulk-like, it is drawn with a dashed line.}
\label{fig:toy2d_3band_surf}
\end{figure}
%%%%%%%%%%%%%%%%%%%%%%%%%%%%%%%%%%%%%%%%%%%%%%%%%%%%%%%%%%

Finally, we study another 2D toy model which manifests the vitalness of the conditions for the bulk-boundary correspondence, such as inversion symmetry and insulating terminated system.
We again confirm that the extra charge accumulation in the surface regions is captured by the inter-cellular Zak phase. 

The toy model-II, depicted in Fig. \ref{fig:toy2d_3band}(a), has a tight-binding model given by
\begin{eqnarray}
\mathcal{H}_{\mathrm{2D}}^{\mathrm{II}} &=& \sum_{\mathbf{R}} \mathbf{c}^\dag_{\mathbf{R}} H_0\mathbf{c}_{\mathbf{R}} + \sum_{\mathbf{R},\boldsymbol\delta_x} \mathbf{c}^\dag_{\mathbf{R}+\boldsymbol\delta_x}H_x\mathbf{c}_{\mathbf{R}} \nonumber\\
&& +\sum_{\mathbf{R},\boldsymbol\delta_y} \mathbf{c}^\dag_{\mathbf{R}+\boldsymbol\delta_y}H_y\mathbf{c}_{\mathbf{R}}
\end{eqnarray}
where $\mathbf{c}_\mathbf{R}^\dag = (a_{1,\mathbf{R}}^\dag,a_{2,\mathbf{R}}^\dag,b_{1,\mathbf{R}}^\dag,b_{2,\mathbf{R}}^\dag,c_{1,\mathbf{R}}^\dag,c_{2,\mathbf{R}}^\dag )$ consists of the creation operators of the orbitals at $A$, $B$ and $C$ sites.
There are two orbitals per site.
The matrices are defined as $H_0 = \mathrm{diag}(\epsilon_{A,1},\epsilon_{A,2},\epsilon_{B,1},\epsilon_{B,2},\epsilon_{C,1},\epsilon_{C,2})$, $H_x = \mathbb{1} \otimes (\sigma^x+\sigma^z)$ and $H_y = (\lambda_1+\lambda_6)\otimes\sigma_z$, where $\mathbb{1}$ is the $3\times 3$ identity matrix, $\sigma^\alpha$ is the Pauli matix and $\lambda_i$ is the Gellman matrix.
The band-structure at $k_y=0$ is depicted in Fig. \ref{fig:toy2d_3band}(b).
While there are four Dirac points at $k_y=0$, as marked by red dots, it is insulating otherwise.
We consider this model because we can investigate a finite system with both reflection symmetry and commensurate bulk's unit cell as shown below.

As in the previous 2D toy model-I, we consider the effective 1D Hamiltonian for a fixed $k_x$ and calculate the \intercellular Zak phases by integrating the Berry connection along the $k_y$ axis.
While $\gamma^\text{inter}$ depends on the choice of the unit cell, we consider the two unit cells, called UC-II(A) and UC-II(S), illustrated in Fig. \ref{fig:toy2d_3band}(a).
Here, we assume that $\epsilon_{A,i} = \epsilon_{C,i}$ so that UC-II(S) preserves reflection symmetry while UC-II(A) does not.
Calculations of $\gamma^\text{inter}$ are shown in Fig.~\ref{fig:toy2d_3zak} by red solid curves along $k_x$.
The \intercellular Zak phase of UC-II(S) is quantized to a multiple of $\pi$ due to the reflection symmetry while it has continuous values for UC-II(A).
For both cases, the \intercellular Zak phase shows the discrete jump at every Dirac node as in the 2D toy model-I.

We consider two terminated systems called Ribbon-II(A) and Ribbon-II(S), which are commensurate with UC-II(A) and UC-II(S) each.
As in the toy model-I, they are terminated along the $x$ axis.
We plot the extra charge accumulations in the bottom surface of those systems in Fig. \ref{fig:toy2d_3zak}.
For the case of Ribbon-II(S), we break the reflection symmetry slightly so that one of the equivalent \intercellular Zak phases $\pi$ and $-\pi$ is selected to describe the extra charge accumulation in the bottom surface.
One can note that the extra charge accumulation is well described by the \intercellular Zak phase for all values of $k_x$ except those in the shaded interval.
This failure in this region is because the condition that the finite system is also insulating is violated.
As shown in Fig. \ref{fig:toy2d_3band_surf}(b), in the shaded region, the topmost band at half-filling joins the conduction band continuum.
Here, `half-filling' means the effective 1D system at each $k_x$ is half-filled.
Also, in this region, the states on the topmost band are no longer exponentially localized.
On the other hand, in Ribbon-II(S), one can see the extra charge accumulation perfectly agrees with the \intercellular Zak phase everywhere.
This is because, in this case, the topmost band at half-filling belongs to the valence band continuum or is located in the middle of the bulk gap and shows a finite gap with empty upper levels as exhibited in \ref{fig:toy2d_3band_surf}(d).
While the correspondence between the inter-cellular Zak phase and the extra charge accumulation is inaccurate only around the Dirac nodes where the system is metallic, one can reduce this inaccuracy to any desired value by increasing the surface region.

Finally, applying the modified bulk-boundary correspondence of Sec. \ref{sec:bulk_boundary}, one can predict the number of surface bands in the gap (even or odd) in the inversion symmetric case.
From the \intercellular Zak phase results in Fig. \ref{fig:toy2d_3zak}(a), one expects an even number of surface bands for $\gamma^\mathrm{inter} = 0$ and an odd number of them for $\gamma^\mathrm{inter} = \pi$ below the Fermi level of the effective 1D system at each $k_x$, except in the shaded region.
As shown in \ref{fig:toy2d_3band_surf}(b), the number of surface modes below the Fermi level is two for $k_x < 1.468$ and zero for $k_x > 2.0944$, which correspond to the intervals where the \intercellular Zak phase is zero.
On the other hand, there is only one filled surface mode in $1.468 < k_x < 2.0944$, except in the shaded part.

\section{Conclusions}
\label{sec:conclusions}

In this work, we have demonstrated that the inter-cellular Zak phase $\gamma^{\mathrm{inter}}$ can predict whether the number of surface modes below the Fermi level in 1D insulators is even or odd, when the commensurate bulk unit cell respects inversion symmetry.
While the Zak phase itself cannot do this due to its arbitrariness depending on the choice of the real-space origin and the unit cell, we have shown that $\gamma^{\mathrm{inter}}$, as an origin-independent quantity, can be exploited for this bulk-boundary correspondence.
Although $\gamma^{\mathrm{inter}}$ also depends on the unit cell choice it is not arbitrary once we select a unit cell that is commensurate with the finite system.
Our bulk-boundary correspondence using $\gamma^{\mathrm{inter}}$ was justified with a microscopic interpretation of $\gamma^{\mathrm{intra}}$ and $\gamma^{\mathrm{inter}}$.
We explicitly showed that $\gamma^{\mathrm{intra}}$ is the electronic part of the bulk dipole moment of the unit cell, and $\gamma^{\mathrm{inter}}$ represents how much weight of the Wannier functions are exchanged with respect to a unit cell boundary.
When the system is terminated, $\gamma^{\mathrm{intra}}$ is interpreted as the classical bound surface charge, while $\gamma^{\mathrm{inter}}$ is understood as the extra charge accumulation around surfaces.
Since the number of surface modes is closely related to the extra charge accumulation, we argue how it is related to $\gamma^{\mathrm{inter}}$ when the commensurate unit cell preserves inversion symmetry.
If the origin is at the inversion center, $\gamma^{\mathrm{inter}}$ becomes identical to the Zak phase, and our bulk-boundary correspondence reduces to the conventional one.
Thereby, our work also clarifies the conditions under which the conventional bulk-boundary correspondence using the Zak phase works.

We expect that the extra charge accumulation can be measured by scanning quantum dot microscopy (SQDM).\cite{Tautz2015}
SQDM offers three-dimensional images of electrostatic potentials down to the subnanometer level from which one could infer the total amount of its source charge.
Since the electric field caused by extra charge accumulations at opposite edges of a long enough 1D chain can be considered independent, the extra charge accumulation at one edge can be obtained from the local electrostatic potential profile.
Therefore, SQDM could be the characterizing experiment for the inter-cellular Zak phase like the capacitance measurement for the Zak phase.

\acknowledgements
We thank D. Vanderbilt, L. Balents, J. E. Moore, F. de Juan, A. G. Grushin and S. Park for useful discussions. This work was supported by the ERC Starting Grant No.\ 679722.

\appendix
\section{Intra- and inter-cellular Zak phases}\label{app:zak phase}

When we calculate the Zak phase, the momentum derivative operates on the exponential factor $e^{ik(ma-x)}$ and the coefficient $\alpha^{n,i,\zeta}_k$ of the lattice periodic part of the Bloch function (\ref{eq:periodic bloch}).
That is,
\begin{align}
\gamma_n =& \im \int_{\mathrm{BZ}} dk \langle u_{n,k}|\partial_k u_{n,k}\rangle \\
=&\im \int_{\mathrm{BZ}} dk \langle u_{n,k}| \sqrt{N}\left(\partial_k e^{ik(ma-x)}\right) e^{-ikma}\psi_{n,k} \rangle  \nonumber \\
& + \im \int_{\mathrm{BZ}} dk \langle u_{n,k}| \sqrt{N} e^{ik(ma-x)} \left(\partial_k e^{-ikma}\psi_{n,k} \right) \rangle
\end{align}
The first term is defined as the intra-cellular Zak phase and the second as the inter-cellular Zak phase.

First, the intra-cellular Zak phase is evaluated as follows.
\begin{widetext}
\begin{eqnarray}
\gamma_n^{\mathrm{intra}} &=& \im\int_{\mathrm{BZ}} dk\int_\Omega dx \sum_{m,m^\prime}^N\sum_{i,i^\prime =1}^{N_b}\sum_{\zeta,\zeta^\prime =1}^{N_{orb}^i} \alpha_k^{n,i,\zeta*}\phi_m^{i,\zeta}(x)^* e^{-\im k(ma-x)} \alpha_k^{n,i^\prime,\zeta^\prime}\phi_{m^\prime}^{i^\prime,\zeta^\prime}(x) \frac{\partial e^{\im k(m^\prime a-x)}}{\partial k}\\
&=& \int_{\mathrm{BZ}} dk \int_\Omega dx \sum_{m,m^\prime}^N\sum_{i,i^\prime =1}^{N_b}\sum_{\zeta,\zeta^\prime =1}^{N_{orb}^i} (x-m^\prime a) \alpha_k^{n,i,\zeta*}\alpha_k^{n,i^\prime,\zeta^\prime} \phi_m^{i,\zeta}(x)^*\phi_{m^\prime}^{i^\prime,\zeta^\prime}(x) e^{\im k(m^\prime - m)a} \label{eq:appendix_1} \\
&=& \int_{\mathrm{BZ}} dk \int_\Omega dx~x \sum_{m,m^\prime}^N\sum_{i,i^\prime =1}^{N_b}\sum_{\zeta,\zeta^\prime =1}^{N_{orb}^i} \alpha_k^{n,i,\zeta*}\alpha_k^{n,i^\prime,\zeta^\prime} \phi_m^{i,\zeta}(x)^*\phi_{m^\prime}^{i^\prime,\zeta^\prime}(x) e^{\im k(m^\prime - m)a} -m_\Omega a\int_{\mathrm{BZ}} dk \sum_{i=1}^{N_b}\sum_{\zeta=1}^{N_{orb}^i}|\alpha_k^{n,i,\zeta}|^2  ~~ \label{eq:appendix_2} \\
&=& N\int_{\mathrm{BZ}} dk \int_\Omega dx~x \left| \frac{1}{\sqrt{N}} \sum_{m}^N\sum_{i=1}^{N_b}\sum_{\zeta=1}^{N_{orb}^i} \alpha_k^{n,i,\zeta} \phi_m^{i,\zeta}(x) e^{\im kma} \right|^2 -2\pi m_\Omega  \label{eq:appendix_3}\\
&=& N\int_{\mathrm{BZ}} dk \int_\Omega dx~x \left| \psi_{n,k}(x) \right|^2 -2\pi m_\Omega 
\end{eqnarray}
\end{widetext}
where $m_\Omega$ is the index of the unit cell $\Omega$.
From (\ref{eq:appendix_1}) to (\ref{eq:appendix_2}), we used the orthonormality condition of the L\"owdin functions, $\int_\Omega dx\phi_m^{i,\zeta}(x)^*\phi_{m^\prime}^{i^\prime,\zeta^\prime}(x) = \delta_{m,m_\Omega}\delta_{m^\prime,m_\Omega}\delta_{i,i^\prime}\delta_{\zeta,\zeta^\prime}$.
The second term of (\ref{eq:appendix_3}) is obtained because the coefficients of the eigenstate for every $n$ and $k$ are normalized.

Second, the inter-cellular Zak phase is calculated as
\begin{widetext}
\begin{eqnarray}
\gamma_n^{\mathrm{inter}} &=& \im\int_{\mathrm{BZ}} dk\int_\Omega dx \sum_{m,m^\prime}^N\sum_{i,i^\prime =1}^{N_b}\sum_{\zeta,\zeta^\prime =1}^{N_{orb}^i} \alpha_k^{n,i,\zeta*}\phi_m^{i,\zeta}(x)^* e^{-\im k(ma-x)} \frac{\partial \alpha_k^{n,i^\prime,\zeta^\prime}}{\partial k}\phi_{m^\prime}^{i^\prime,\zeta^\prime}(x) e^{\im k(m^\prime a-x)} \label{eq:appendix A6}\\
&=& \im\int_{\mathrm{BZ}} dk \sum_{m,m^\prime}^N\sum_{i,i^\prime =1}^{N_b}\sum_{\zeta,\zeta^\prime =1}^{N_{orb}^i} \alpha_k^{n,i,\zeta*} e^{-\im kma} \frac{\partial \alpha_k^{n,i^\prime,\zeta^\prime}}{\partial k} e^{\im k m^\prime a} \delta_{m,m_\Omega}\delta_{m^\prime,m_\Omega} \delta_{i,i^\prime}\delta_{\zeta,\zeta^\prime}\label{eq:appendix A7} \\
&=& \im\sum_{i=1}^{N_b}\sum_{\zeta =1}^{N_{orb}^i} \int_{\mathrm{BZ}} dk \alpha^{n,i,\zeta *}_k \frac{\partial}{\partial k}\alpha^{n,i,\zeta}_k.\label{eq:inter zak 9}
\end{eqnarray}
\end{widetext}
The inter-cellular Zak phase is represented by the Wannier coefficients $A_m^{n,i,\zeta}$ by using the inverse transformation of (\ref{eq:wannier_amp}), which is given by
\begin{eqnarray}
\alpha_k^{n,i,\zeta} = \sum_m^N A_m^{n,i,\zeta} e^{-\im kma}.
\end{eqnarray}
Substituting into Eq.~\eqref{eq:inter zak 9}, we have
\begin{eqnarray}
\gamma_n^{\mathrm{inter}} &=& \sum_{i,\zeta}\sum_{m,m^\prime} \int_{\mathrm{BZ}} dk~m A_m^{n,i,\zeta *}A_{m^\prime}^{n,i,\zeta} e^{\im k(m-m^\prime)a}\quad\quad \label{eq:inter zak 11}\\
&=& 2\pi \sum_{m}\sum_i^{N_\mathrm{b}}\sum_\zeta^{N_{\mathrm{orb}}^i} m \left| A^{n,i,\zeta}_{m} \right|^2,
\end{eqnarray} 
where the summation ranges of $i$ and $\zeta$ are the same as those of (\ref{eq:inter zak 9}).

We further split $\gamma_n^{\mathrm{inter}}$ into $\gamma^{R\rightarrow L}_n$ and $\gamma^{R\rightarrow L}_n$, which satisfies $\gamma_n^{\mathrm{inter}} = -\gamma^{R\rightarrow L}_n + \gamma^{R\rightarrow L}_n$, as follows.
\begin{eqnarray}
\gamma^{R\rightarrow L}_n &=& -2\pi \sum_{m=-\infty}^{-1}\sum_{i=1}^{N_\mathrm{b}}\sum_{\zeta=1}^{N_\mathrm{orb}^i} m \left| A^{n,i,\zeta}_{m} \right|^2 \\
&=& 2\pi \sum_{m^\prime = 0}^\infty \sum_{m=-\infty}^{-1}\sum_{i=1}^{N_\mathrm{b}}\sum_{\zeta=1}^{N_\mathrm{orb}^i} \left| A^{n,i,\zeta}_{m-m^\prime} \right|^2 \label{eq:gamma rl 1}\\
&=& 2\pi \sum_{m^\prime = 0}^\infty \int_{-\infty}^{x_b} dx \left| W_{n,m^\prime}(x)\right|^2 
\end{eqnarray}
and
\begin{eqnarray}
\gamma^{L\rightarrow R}_n &=& 2\pi \sum_{m=0}^{\infty}\sum_{i=1}^{N_\mathrm{b}}\sum_{\zeta=1}^{N_\mathrm{orb}^i} m \left| A^{n,i,\zeta}_{m} \right|^2 \\
&=& 2\pi \sum_{m^\prime = -\infty}^{-1}\sum_{m=0}^{\infty}\sum_{i=1}^{N_\mathrm{b}}\sum_{\zeta=1}^{N_\mathrm{orb}^i} \left| A^{n,i,\zeta}_{m-m^\prime} \right|^2 \label{eq:gamma rl 2}\\
&=& 2\pi \sum_{m^\prime=-\infty}^{-1} \int^{\infty}_{x_b} dx \left| W_{n,m^\prime}(x)\right|^2 
\end{eqnarray}
where $x_b$ is the boundary between the $m=-1$ and the $m=0$ unit cells.
Although we specify the boundary $x_b$, the result is independent of it due to translational invariance, as it is clear from the dependence on $m-m^\prime$ in Eqs.~\eqref{eq:gamma rl 1} and~\eqref{eq:gamma rl 2}.

\section{Bound surface charge}\label{app:bound charge}

In this section, we show that the bound surface charge can be split into the classical bound surface charge and the extra charge accumulation.
The bound surface charge of the left edge is given by
\begin{align}
\sigma^{LS}= \frac{1}{a}\int_{-\infty}^{x_c}dx \int_{x - \frac{a}{2}}^{x + \frac{a}{2}}dx^\prime \rho_{\text{t.s.}}(x^\prime) =\int_{-\infty}^{x_c}dx \bar{\rho}_{\text{t.s.}}(x)
\end{align}
where $\rho_{\text{t.s.}}(x)$ is the total charge density including both the electronic and ionic contributions, $\bar{\rho}_{\text{t.s.}}(x) = a^{-1}\int_{x - \frac{a}{2}}^{x + \frac{a}{2}}dx^\prime \rho_{\text{t.s.}}(x^\prime)$, and $x_c$ is an arbitrary position in the finite system, far away from the surfaces compared with the widths of the $\varphi_l^{LS}$'s.

First, note that $\sigma^{LS}$ is independent of $x_c$ since
\begin{eqnarray}
\frac{\partial \sigma^{LS}}{\partial x_c} = \frac{1}{a} \int_{x_c - \frac{a}{2}}^{x_c + \frac{a}{2}}dx^\prime \rho_{\text{t.s.}}(x) = \bar{\rho}_{\text{t.s.}}(x_c)
\end{eqnarray}
vanishes when $x_c$ is far from the surfaces.
The integral of the charge density over one unit cell length $a$ is zero far from the surfaces because our system is assumed to be neutral.

Second, the expression for the bound surface charge can transformed as follows:
\begin{eqnarray}
\sigma^{LS} &=& -\int_{-\infty}^{x_c}dx \frac{d \bar{\rho}_{\text{t.s.}}(x)}{dx}x + \bar{\rho}_{\text{t.s.}}(x)x \Big|_{-\infty}^{x_c} \label{eq:sigma_L 1}\\
&=& -\frac{1}{a}\int_{-\infty}^{x_c}dx \left[ \rho_{\text{t.s.}}(x+\frac{a}{2}) - \rho_{\text{t.s.}}(x-\frac{a}{2})\right]x ~~~\label{eq:sigma_L 2}\\
&=& -\frac{1}{a}\int_{x_c-\frac{a}{2}}^{x_c+\frac{a}{2}}dx \rho_{\text{t.s.}}(x)x + \int_{0}^{x_c-\frac{a}{2}}dx\rho_{\text{t.s.}}(x) \label{eq:sigma_L 4}
\end{eqnarray}
where $x_\pm = x \pm a/2$.
We have used the fact that $\bar{\rho}_{\text{t.s.}}(x_c) = \bar{\rho}_{\text{t.s.}}(-\infty) =0$ in going from \eqref{eq:sigma_L 1} to \eqref{eq:sigma_L 2}, and that $\rho_{\text{t.s.}}(x)=0$ for $x<0$ from \eqref{eq:sigma_L 2} to \eqref{eq:sigma_L 4}.
If $x_c = x_{\ell_L}+a/2$, we have
\begin{align}
\sigma^{LS} =-\frac{1}{a}\int^{x_{\ell_L}+a}_{x_{\ell_L}}dx~x\rho_{\text{t.s.}}(x) + \int_{0}^{x_{\ell_L}}dx\rho_{\text{t.s.}}(x).
\end{align}
In the same way, the bound surface charge at the right edge becomes
\begin{align}
\sigma^{RS} = \frac{1}{a}\int^{x_{N-\ell_R}}_{x_{N-\ell_R}-a}dx~x\rho_{\text{t.s.}}(x) + \int^{x_{N}}_{x_{N-\ell_R}}dx\rho_{\text{t.s.}}(x).
\end{align}
These are the expressions for the bound surface charges used in (\ref{eq:left bound charge}) and (\ref{eq:right bound charge}) in Sec.~\ref{sec:extra_charge}.

%\bibliography{sample}

\end{document}